\documentclass[twocolumn,showpacs,preprintnumbers,amsmath,amssymb]{revtex4}
\usepackage{amsfonts,amssymb,graphicx}
\def\be{\begin{equation}}
\def\ee{\end{equation}}
\def\bea{\begin{eqnarray}}
\def\eea{\end{eqnarray}}
\def\ba{\begin{eqnarray*}}
\def\ea{\end{eqnarray*}}
\def\<{\langle}
\def\>{\rangle}
\def\~{\tilde}
\def\s{\sigma}

\def\t{\tau}

\begin{document}
\title{Factorization Properties in the 3D Edwards-Anderson Model}
\author{Pierluigi Contucci}
\email{contucci@dm.unibo.it}
\affiliation{Dipartimento di Matematica, Universit\`{a} di Bologna\\
Piazza di Porta S.Donato 5, 40127 Bologna, Italy}
\author{Cristian Giardin\`a}
\email{giardina@eurandom.tue.nl}
\affiliation{EURANDOM \\
P.O. Box 513 - 5600 MB Eindhoven, The Netherlands}
\begin{abstract}
\noindent
Starting from the study of a linear combination of multi-overlaps which can be
rigorously shown to vanish for large systems we numerically analyze the factorization
properties of the link-overlaps multi-distribution for the $3D$ Gaussian Edward-Anderson
spin-glass model. We find evidence of a pure factorization law for the multi-correlation
functions.
For instance the quantity
$$
\frac{<Q_{12}^2> - <Q_{12}Q_{34}>}{<Q_{12}^2>}
$$
tends to zero at increasing volumes.
We also perform the same analysis
for the standard overlap for which instead the lack of factorization persists increasing the
size of the system. The necessity of a better understanding of the mutual relation between
the two overlaps is pointed out.
\end{abstract}
\maketitle
The structure of the low temperature phase for the finite dimensional spin glass
is among the most interesting and yet unsettled problems in condensed matter.
After more than thirty years its main issues remain unsolved not only from
the mathematically rigorous point of view but also within the theoretical
physics perspective. In particular it is not clear what is the quenched probability
distribution of the spin-overlap for large systems and low temperatures and different
pictures have been proposed: the Replica Symmetry Breaking \cite{MPV} which describes the
distribution similarly to the mean field one, the Droplet picture \cite{FH}
which claims that the support of the distribution shrinks to a point (becomes trivial) when the
system size grows to infinity (see also \cite{NS,KM,PY} for other possible pictures
and different perspectives).

In this paper we investigate the structure of the multi-distribution of the
overlaps among replicated samples of the Edwards-Anderson
(EA) model with Gaussian nearest neighbor couplings. The spin glass quenched measure
is indeed described (see \cite{G,C} for instance) by an infinite family of probability
distributions which represent its equilibrium state: the distribution of the single
overlap $P({\cal Q}_{1,2})$ related to the internal energy of the system but also those
involving more than two replicas like $P({\cal Q}_{1,2},{\cal Q}_{1,3})$, $P({\cal Q}_{1,2},{\cal Q}_{3,4})$ related
to the specific heat etc. Since in the quenched measure the different copies are
taken with the same frozen disorder the random variables ${\cal Q}_{l,m}$ are not independent
and their joint distribution does not factorize on products of the single one at finite
volumes: $P({\cal Q}_{1,2},{\cal Q}_{1,3})\neq P({\cal Q}_{1,2})P({\cal Q}_{1,3})$.

With this work we address precisely the question if such a factorization
may occur when the thermodynamic limit is reached and we find strong numerical evidence
for a positive answer in the link overlap case and a negative one for the standard overlap.

Previous studies on the model \cite{KM,PY}
have concentrated their attention on the single overlap distribution and claims were made
about its triviality (for different perspectives see also \cite{MP2,MP,NS2}). We independently
reproduced those numerical results and extended them to larger sizes (up to $L=12$).
In our opinion they cannot distinguish among the trivial or non-trivial picture. In particular
the data for the link overlap do not rule out a limiting distribution with two peaks and a
plateau among them.
That fact together with a complete lack of rigorous results make the factorization
properties of the multi-overlaps a completely open matter and motivate our investigation.

Our departing point are the factorization rules found for the Edwards-Anderson model
in terms of its link-overlap. In a box of side $L$ the {\em link overlap} between two
spin configuration $\sigma$ and $\tau$ is defined as
\begin{equation}
\label{qlink}
Q_{L}(\sigma, \tau) \, = \, \frac{1}{3L^3}\sum_{(i,j)}\s_i\s_j\t_i\t_j
\end{equation}
where the sum runs over all couples of sites which are connected by
a random bond (usually the nearest neighbor sites).
In \cite{C} and \cite{CG} it was rigorously proved
that when the size of the system grows to infinity the
linear combination
\begin{equation}
<Q_{1,2}^2-4Q_{1,2}Q_{2,3}+3Q_{1,2}Q_{3,4}>
\label{lac}
\end{equation}
is vanishing except possibly on isolated temperatures where phase transitions may
occur. Here the brackets denote the
quenched measure (i.e. the successive computation of the expectation w.r.t. the
Boltznmann-Gibbs measure on replicated samples followed by the average over the
Gaussian disorder) and the subscripts on the $Q$'s label the different real
replicas.

The same relation (\ref{lac}) was originally found to hold
for the square of the {\em standard overlap}
\begin{equation}
\label{qstan}
q^2_{L}(\sigma, \tau) \, = \, \left(\frac{1}{L^3}\sum_{i}\s_i\t_i\right)^2
\end{equation}
in the SK model within the RSB picture
and related to a property of {\it replica equivalence} \cite{AC,P}.
See also \cite{G,GG,CG} for its rigorous derivations.

Although the vanishing of (\ref{lac}) emerged first in the mean field picture
its validity alone cannot distinguish between the different scenarios proposed
(see the discussion in \cite{C}).
In particular it cannot distinguish between the peculiar ultrametric factorization rule
like the one proposed for the SK model or a pure factorization rule of the multi-overlap
distributions.

In order to test the factorization properties we chose to
consider generic linear combinations of the above monomials with coefficients
whose sum is zero:
\begin{equation}
g(\alpha) \, = \, < Q_{1,2}^2-\alpha Q_{1,2}Q_{2,3} + (\alpha-1) Q_{1,2}Q_{3,4} > \; .
\label{alpha}
\end{equation}
From \cite{CG} we know that the former expression is close to zero in $\alpha=4$
with finite volume correction of size $L^{-3}$. Away from $\alpha=4$ a
link-overlap multi-distribution with a pure factorization law
between replicas would predict a progressive squeezing to zero (at increasing volumes)
of the line $g(\alpha)$ for all $\alpha$ .
Instead a non-factorizing link-overlaps distribution
would be compatible with the persistence away from zero
of the line angular coefficient
\be
\label{m}
m = <Q_{1,2}Q_{3,4}- Q_{1,2}Q_{2,3}>
\ee
and intercept
\be
\label{n}
n = <Q_{1,2}^2-Q_{1,2}Q_{3,4}>.
\ee

\begin{table}
\begin{tabular}{|c||c|c|c|c|c|c|c|} \hline
$L$ & Therm. & Equil. & Samples & $N_{\beta}$ & $\delta T$ &
$T_{min}$& $T_{max}$\\ \hline \hline
3    & 50000  &  50000    & 2048   &  19  &  0.1   & 0.5 & 2.3 \\ \hline
4    & 50000  &  50000    & 2048   &  19  &  0.1   & 0.5 & 2.3 \\ \hline
6    & 50000  &  50000    & 2048   &  19  &  0.1   & 0.5 & 2.3 \\ \hline
8    & 50000  &  50000    & 2680   &  19  &  0.1   & 0.5 & 2.3 \\ \hline
10   & 70000  &  70000    & 2050   &  37  &  0.05  & 0.5 & 2.3 \\ \hline
12   & 70000  &  70000    & 2032   &  37  &  0.05  & 0.5 & 2.3 \\ \hline
\end{tabular}
\caption{Parameters of the simulations:
system size, number of sweeps used for
thermalization, number of sweeps during which
observables were measured, number of disorder samples,
number of $\beta$ values allowed in the parallel tempering
procedure, temperature increment,
minimum and maximum temperature values.}
\label{parameter}
\end{table}

We performed the analysis for both the standard overlap and for the link overlap.
Before going to the detailed descriptions of the model and the illustration of the results
it is worth to mention the different roles played
by the two. The standard overlap is historically
the first proposed observable in the study of the spin glass phase being directly related
to the original Edwards-Anderson order parameter. Its widespread use is moreover due
also to the fact that in the SK model its distribution carries the
whole information of the thermodynamic properties. Undoubtedly it is a very interesting
quantity to be studied for the low temperature phase of general disordered spin systems;
nevertheless we want to point out that each spin glass model has its own natural observable
which, in a Gaussian model, is given by the covariance of its Hamiltonian. An easy computation
\cite{C} shows that in the EA model such a covariance is the link-overlap
in terms of which all the thermal observable can be expressed like internal energy, specific heat etc,
and rigorous results can be established like stochastic stability \cite{AC,CG}; see also \cite{NS3}
for its use in rigorous results from a different perspective.
It is also interesting to observe that the standard overlap between two spin
configuration changes proportionally to the volume of the different spins in the two
configurations; instead the link overlap changes like the surface.

Recalling the basic definitions: we consider the Gaussian Edward-Anderson
model \cite{EA}, defined by the Hamiltonian
\be
H(J,\s) = - \sum_{i \in \Lambda} \sum_{\mu = x,y,z}
J_{i,i+e_{\mu}}\s_i\s_{i+e_{\mu}}
\ee
where $i$ is a site of a 3-dimensional cubic lattice $\Lambda$ ($|\Lambda | = L^3$),
$e_{\mu}$ is the versor in the $\mu$ direction (with $\mu = x,y,z$),
$\s_i=\pm 1$  are Ising spin variables and $J_{i,i+e_{\mu}}$
are Gaussian random variables with zero average and unit variance.

We performed numerical simulations by using the
Parallel Tempering (PT) algorithm to facilitate equilibration.
We used periodic boundary conditions and investigated
lattice sizes up to $L=12$. For every size we simulated
at least $2032$ realizations of the couplings.
Other parameters of the simulations are reported in
Table (\ref{parameter}).
The allowed temperature range (assuming $T_c\simeq 1.$)
was approximately $0.5 T_c < T < 2.3 T_c$ and we used up
to $37$ temperatures in the PT procedure.
We tested thermalization by checking the symmetry of the probability
distribution for the standard overlap under the transformation
$q\rightarrow -q$.

\begin{figure*}[h!]
\includegraphics[width=6.5cm,angle=-90]{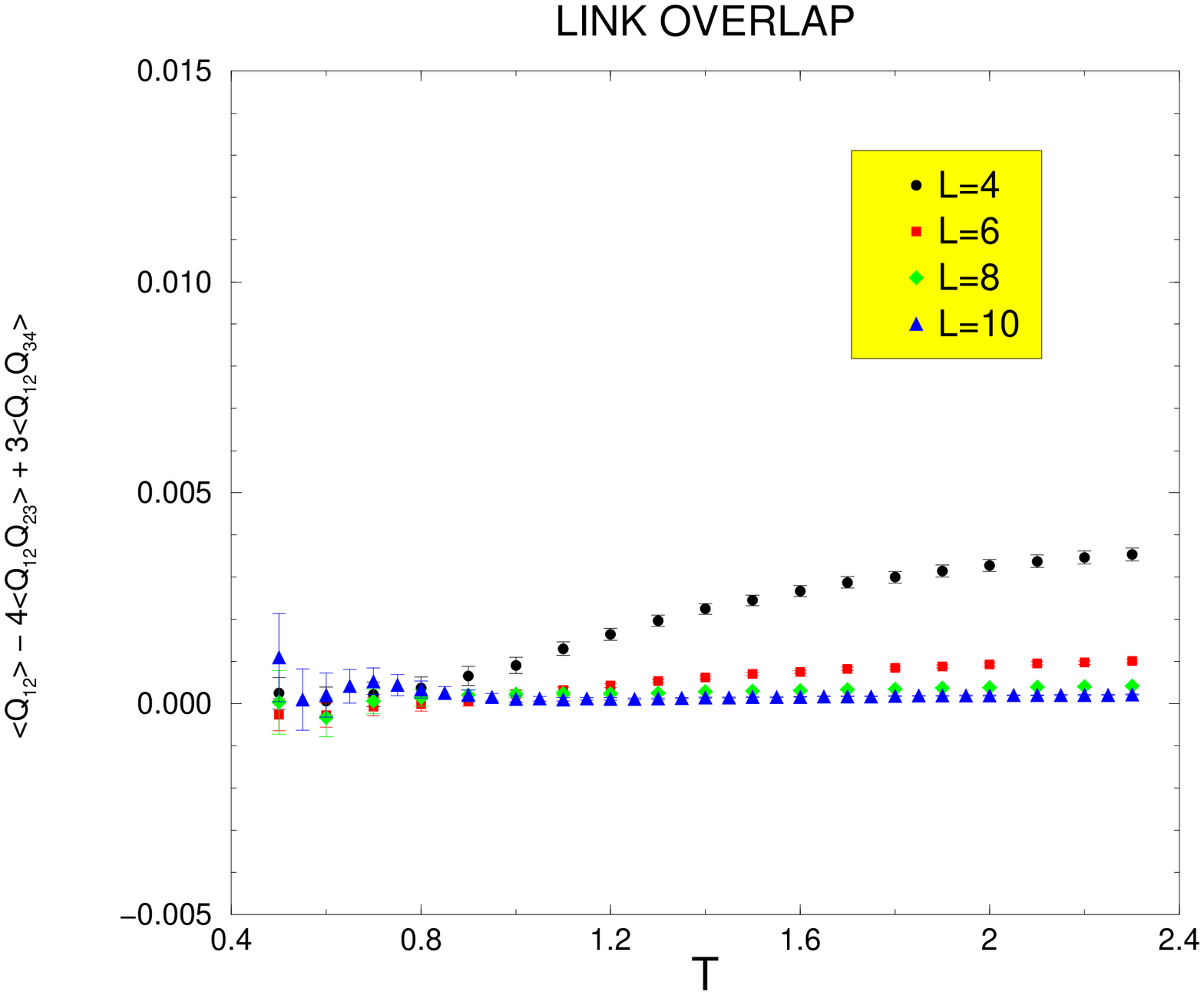}
\hspace{1.0cm}
\includegraphics[width=6.5cm,angle=-90]{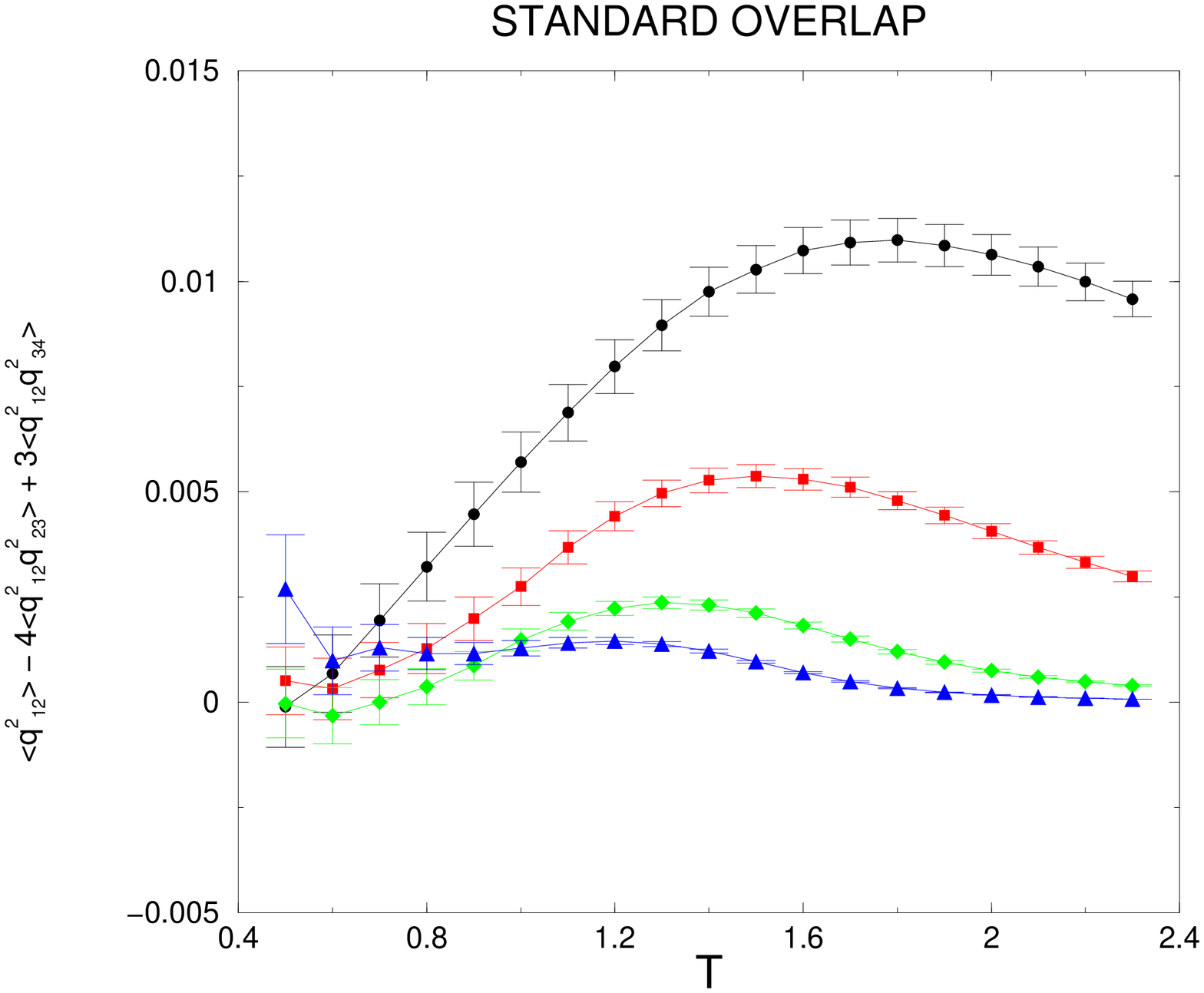}
\caption{\label{fig3} Plot of relation (\ref{lac})
as a function of the temperature for different system sizes
(see the legend). On the left data for the link overlap,
on the right data for the standard overlap.}
\end{figure*}

If not otherwise stated in the sequel we always plot the same
quantities relative to the two overlap
with the same scale on the $y$-axis, in order to let better
appreciate analogies and differences between  the two.

Fig. (\ref{fig3}) shows the plot of relation in Eq. (\ref{lac}) as a function
of the temperature; we see on the left side that the vanishing of the linear
combination is well reproduced by numerical simulation for the link overlap
(even for the small size $L=4$  the finite size correction is less than
$5\cdot 10^{-3}$). It is interesting to see that also for the squared
standard overlap (right side) the relation is satisfied (in agreement
with \cite{MPRRZ}), even if there are no rigorous argument which support it.
We notice moreover that finite size corrections are larger than those for the link overlap.

\begin{figure*}[h!]
\includegraphics[width=6.cm,angle=-90]{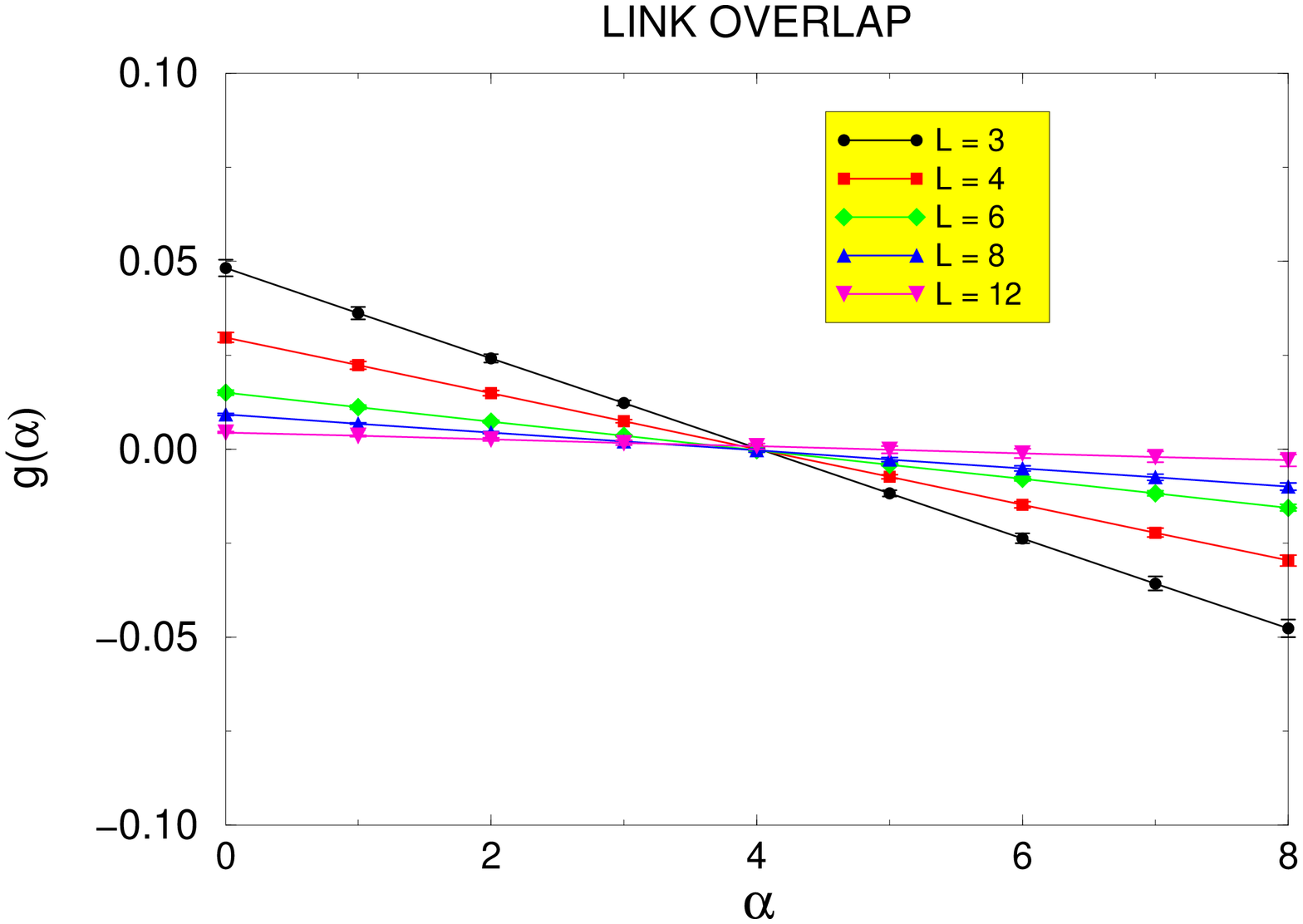}
\hspace{1.cm}
\includegraphics[width=6.cm,angle=-90]{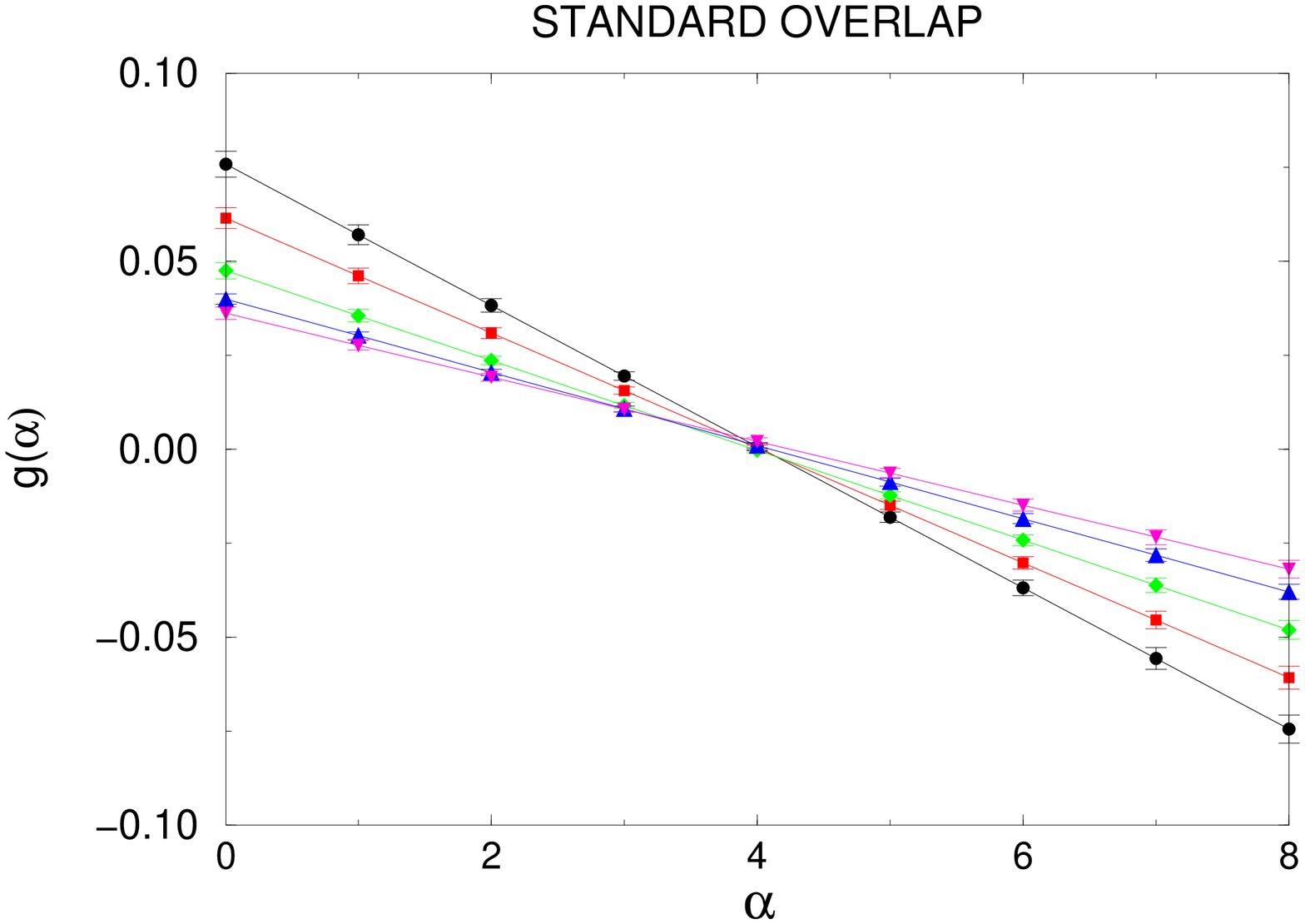}
\caption{Plot of the function $g(\alpha)$ versus $\alpha$
for the fixed temperature $T=0.6$: link overlap(left), standard overlap (right).}
\label{fig5}
\end{figure*}

In Fig. (\ref{fig5}) we show the different lines $g(\alpha)$
as the volume increases.
For all the sizes the relation is close to zero
for $\alpha = 4$. When the system size increases we observe that
the lines tend to flatten, which is a possible signal of triviality.
We may observe that the flattening is much more evident for the link overlap
compared to the standard overlap.

To investigate further the factorization and triviality matter, we measure the angular
coefficient (\ref{m}) and the intercept (\ref{n})
and see how they scale w.r.t. to the system size $L$. We find that both have some
tendency to decrease and we fit the data with a law of the type
$
y = a + c\cdot L^{\delta}
$
for different values of $a$ and measuring the relative chi-square. We find that
the minimum chi-square for the angular coefficient
and intercept relative to the link-overlap
is reached for $a=0$ which supports the factorizing picture. More precisely
the data gives a normalized $\chi^2 = 1.5$ for $a=0$ and the $\chi^2$ value
is increasing with $a$, being already of ${\cal O} (10)$ for $a=0.001$.
On the other hand the same analysis for the standard overlap showed that the chi-square is
basically constant for all values of $a$ close to zero.
It has a normalized $\chi^2 = {\cal O}(1)$ for all values in the interval
$[0,0.001]$. These results suggests that the data we found for the standard overlap do not allow
to distinguish among a factorizing picture ($a=0$ in the thermodynamic limit) and a \
non-factorizing ($a\neq 0$).

\begin{figure}[h!]
\includegraphics[width=6.cm,angle=-90]{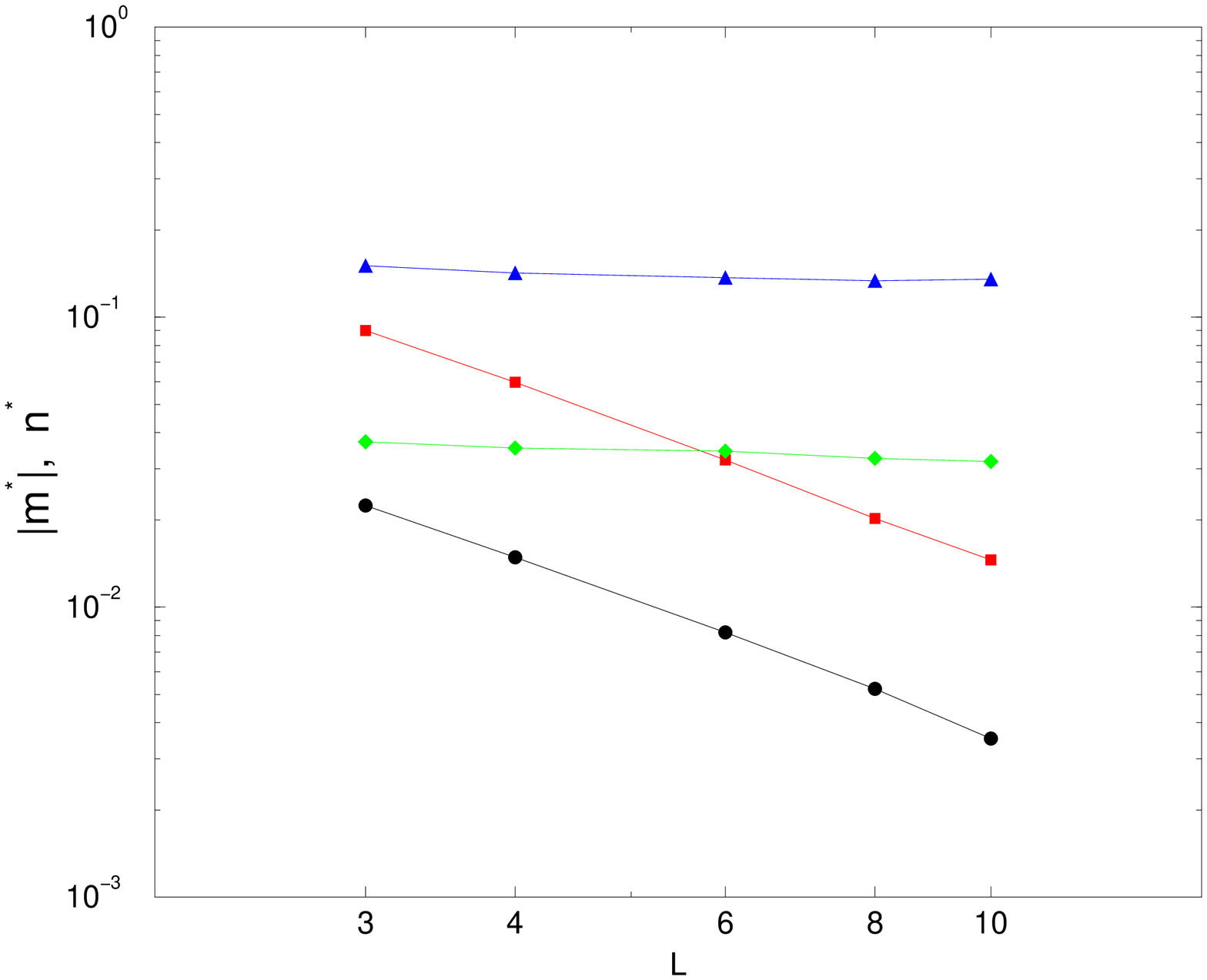}
\caption{The absolute value of the normalized angular coefficient $m^{*}$
(black circles for the link overlap, green diamond for standard overlap)
and intercept $n^{*}$ (red squares for the link overlap, blue triangles
for the standard overlap) versus the lattice size $L$. The temperature is
$T=0.6$. Lines are guide to the eyes.}
\label{fig9}
\end{figure}

Finally we analyzed the normalized angular coefficient
$m^{*} = m/<Q_{12}^2>$ and intercept $n^{*} = n/<Q_{12}^2>$
which are adimensional quantities normalized with their typical values
(they are slightly different for the two overlaps).
The data reported in log-log scale in Fig.(\ref{fig9}) show a clear
factorization tendency for the link overlap multi-distribution while
the standard overlap seem to lack the same property. The chi-square analysis
confirms all that.
This works shows that, within the current reachable lattice sizes,
the multi distribution for the link overlap obeys to a pure factorization
law while the standard overlap lacks the same property. A similar factorization
has a very clear meaning: the random variables $Q_{l,m}$ become independent in the
infinite volume limit with respect to the quenched measure. This fact is certainly
a form of triviality and is observed for instance in the Curie-Weiss model
where the spin variables $\s_i\s_j$ become independent in the infinite volume limit
with respect to the Boltzmann measure (see \cite{CGI}). We want to stress nevertheless
that the factorization triviality that we found is compatible with both the usual trivial
or nontrivial pictures because it doesn't imply anything on the single overlap distribution
but only on the multi overlap ones.

\begin{figure}[h!]
\includegraphics[width=7.5cm,angle=-90]{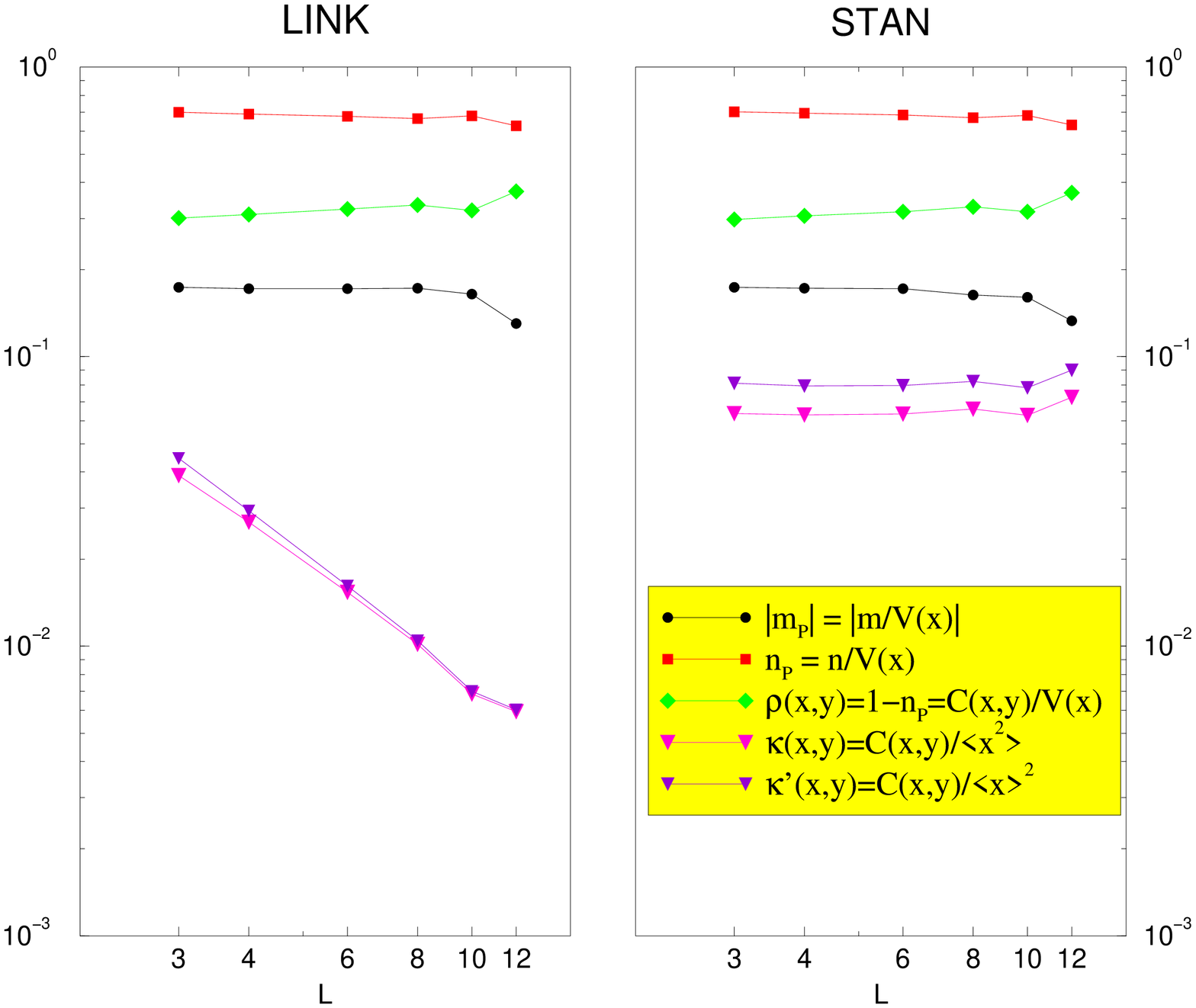}
\caption{A different normalization of the data (see the legend).
Here $x=Q_{12}$, $y=Q_{34}$ (left panel) and
$x=q_{12}^2$, $y=q_{34}^2$, (right panel),
$V(x)=\<x^2\>-\<x\>^2=V(y)$ denotes the variance and
$C(x,y) = \<xy\> -\<x\>\<y\>$ denotes the covariance.}
\label{norm-final}
\end{figure}

We observe, finally, that the different behavior of the two overlaps requires a better understanding
of their relative fluctuations: does the standard overlap distribution concentrate at
fixed link overlap? And if it does is the functional relation among the two a
one-to-one function? The two questions can be trivially answered in the SK model
where the link overlap is the square power of the standard overlap.
In a finite dimensional model like EA the relation among the two is much more
involved and the problem of their relative fluctuation has been numerically
addressed in \cite{MP} at $T=0$ and more recently in \cite{S} for a model
with soft constraints. Their results point to the direction of a
vanishing relative fluctuation.
The results we found on this paper would be compatible either with a non vanishing
relative fluctuation or with a multivalued functional relation among the two random
variables that could account for the difference in factorization behavior.
We plan to investigate further those two open questions both numerically and analytically.
\footnote{After the first appearance of this work in the arxiv
(March 7th 2005) Giorgio Parisi suggested to us to investigate a different
normalization of the parameters $n$ and $m$ with the variance
of the distribution at the denominator, we will denote them $n_P$ and $m_P$. We extended his
suggestions to all the second order quantities. Some results are shown in Fig.(\ref{norm-final}).
What we observe is the remarkable coincidence of the $n_P$ and $m_P$ for the two different overlaps.
Introducing the correlation coefficient $\rho$ and observing that $n_P=1-\rho$ the identical
behavior of $\rho$ for the two overlaps suggests a high mutual correlation among them.
The fact that $\rho$ stays basically constant in the observed volume range does not
give much information on the factorization of the multi overlap because it may happen
that the variance at the denominator shrinks to zero like the covariance at the numerator.
Instead the coefficient $k$ normalized with the second order moment shows the same behavior
of $m^*$ and $n^*$ and confirms our conclusions.}

We thank S. Graffi, F.Guerra, E. Marinari, C. Newman, M. Palassini, G.Parisi, F. Ricci-Tersenghi, D. Sherrington,
N. Sourlas, D.Stein and F. Unguendoli for interesting discussions.

\end{document}